 \documentclass[preprint2]{aastex}

\slugcomment{Submitted for publication in ApJ. Lett.}

\begin{document}

\title{Low-energy X-ray emission\\
    from the Abell 2199 cluster of galaxies}

\author{Jelle S. Kaastra}
\affil{SRON  National Institute for Space Research,
Sorbonnelaan 2, 3584 CA Utrecht, The Netherlands}
\email{j.kaastra@sron.nl}

\author{Richard Lieu}
\affil{Department of Physics, University of Alabama, Huntsville, AL 35899}

\author{Johan A.M. Bleeker and Rolf Mewe}
\affil{SRON  National Institute for Space Research,
Sorbonnelaan 2, 3584 CA Utrecht, The Netherlands}

\and

\author{Sergio Colafrancesco}
\affil{Osservatorio Astronomico di Roma, Via dell'Osservatorio 2,
I-00040 Monteporzio, Italy}

\begin{abstract}
In a recent Letter, Bergh\"ofer \& Bowyer rediscussed the analysis of BeppoSAX
LECS data of the cluster of galaxies Abell~2199 as presented by Kaastra et al.,
in particular the detection of a soft X-ray excess.  Bergh\"ofer \& Bowyer
stated that their analysis method is better suited and does not show evidence
for a soft X-ray excess.  Here we find it necessary to publish a rebuttal,
because it can be demonstrated that the method used by Bergh\"ofer \& Bowyer is
oversimplified, leading to an erroneous result.  As a consequence, their
statement that our initial analysis is incorrect is invalid and the detection of
a soft X-ray excess in A~2199 is still confirmed.
\end{abstract}


\section{Introduction}

The X-ray spectrum of the cluster of galaxies Abell\,2199 has been studied by
many instruments.  The detection of a soft X-ray excess in this cluster was
first claimed by \citet{bowyer98}, based upon {\it Extreme Ultraviolet Explorer
(EUVE)} data, although they do not represent any analysis work.
\citet{kaastra99} analyzed the {\it BeppoSAX} data of this cluster of galaxies
and found evidence of both a soft and a hard X-ray excess at radii larger than
300\,kpc.  This last analysis was based on spatially resolved spectroscopy with
data from the {\it BeppoSAX}, {\it EUVE} and {\it ROSAT} missions.

However, in a recent Letter, \citet{berghoefer02}, (hereafter BB) made
categorical statements that the analysis of Kaastra et al.  is flawed.  BB say
explicitly that "Unfortunately, the telescope sensitivity profile used is likely
to be incorrect," that "{\it BeppoSAX} LECS does not detect an EUV excess when
the data are analyzed correctly," that "using a procedure better suited to the
analysis of extended sources we show that there is no excess in Abell~2199," and
that "these findings appeared to support the (incorrect) finding of an excess in
this cluster using {\it EUVE} data."  In this Letter, we discuss the recent
conclusions of BB, and we show that BB in fact did not use a different
"telescope sensitivity profile" and that their conclusions are not based upon
improved calibration knowledge.  Therefore, in this Letter, we demonstrate that
by pursuing their own highly nonstandard approach to {\it BeppoSAX} data, the
results of BB are susceptible to many problems.

For a clear discussion of this controversy, we focus on the data analysis
procedures and omit the discussion on the physical implications of the existence
of a soft excess, since this has already been presented in our earlier work
\citep{kaastra99}.

\section{Summary of the data analysis by Kaastra et al.}

The data analysis method used previously by us has been described by
\citet{kaastra99}.  Briefly, the cluster was divided into seven concentric
annuli, centered around the bright cD galaxy, with outer radii of 3\arcmin,
6\arcmin, 9\arcmin, 12\arcmin, 15\arcmin, 18\arcmin, and 24\arcmin.
Background-subtracted spectra for these regions were obtained from the {\it
BeppoSAX} Low and Medium Energy Concentrator Spectrometers (LECS and MECS,
respectively), High Pressure Gas Scintillator Proportional Counter, and Phoswich
Detection System as well as from the {\it ROSAT} Position Sensitive Proportional
Counter (PSPC) and the {\it EUVE} Deep Survey (DS).  The background for the {\it
BeppoSAX} LECS and MECS instruments was obtained from the standard blank-sky
observations, and it was verified that the background level during our
observation was not enhanced due to a higher level of particle activity.

For the study of the soft excess, the most important instruments are the LECS,
PSPC, and DS, but the data of all instruments in the entire 0.1--100~keV range
were used and fitted simultaneously.

The instrumental point-spread function (PSF) of the LECS instrument is a strong
function of energy \citep{parmar97}.  For example, at 0.28~keV, the energy
resolution is 32~\% (FWHM) and the angular resolution 9.7\arcmin\ (FWHM).  Both
scale approximately with the incoming photon energy $E$ as $E^{-0.5}$.
Moreover, the wings of the PSF are strongly non-Gaussian.  Thus, at low
energies, a significant fraction of the flux generated in a given annulus
extends to the neighboring annuli, so that the annular spectra are coupled and
need to be fitted {\it simultaneously}.

Therefore, in our analysis, we fitted the spectra of all eight annuli and all
instruments (26 spectra in total) simultaneously.  In the response matrices, the
position- and energy-dependent vignetting factors were taken into account as
well as the effects of the (energy-dependent) overlapping PSF for the different
annuli.

The spectral model that we used consists of a thermal plasma in collisional
ionization equilibrium for each annulus.  For the inner regions, the possible
effects of resonance scattering have also been taken into account for the iron
K$\alpha$ complex.  In addition, a cooling-flow model with partial absorption
has been included in the central annulus.

After applying this model to the data, we found evidence at the 99.99~\%
confidence level of the soft excess as described by \citet{kaastra99}.  In
particular, in the 6\arcmin--12\arcmin\ range, the 0.1--0.3\,keV excess
luminosity above the thermal component is 25\,\%.  In this region, the
subtracted background is smaller than 15\,\% of the cluster signal, while the
large scale variations of the background in this region of the sky are less than
10\,\% of this background.

\section{Summary of the data analysis by BB}

\subsection{Background subtraction}

BB focus much of their attention on the background subtraction.  However, as we
noted above, this is not really a matter of concern for the observed cluster
signal since (1) the background in the relevant radial range is relatively small
and (2) its level during the A\,2199 observation was of average level and not
enhanced.

BB attempted to divide the background into a flat, time-dependent particle
background plus an X-ray background.  Much earlier, \citet{parmar99a,parmar99b}
already showed that the particle contribution in the 0.1--0.5~keV band is small,
less than 13~\% of the total, and that it even decreased by about 15~\% over two
years (our A\,2199 observation is in the middle of this period).  Also, a study
of the MECS background (see F.  Fiore, D.  Ricci, \& P.  Giommi 1997)\footnote{
BepSDC Report on LECS and MECS Dark Earth Background
(http://bepposax.gsfc.nasa.gov/bepposax/soft\-ware/cookbook/rep\_dark\_497.html)}
indicates that the particle background does not vary in time by more than 30~\%
of the total background.  The remaining background is the cosmic X-ray
background.  Since the LECS only operates during satellite night time, any
contribution to the background from scattered solar X-rays is negligible
\citep{parmar99b}.

However, BB scaled the remaining non-particle background for A\,2199 and A\,1795
by a factor of 4.4 and 3.9, respectively, and justified the magnitude of the
scaling factor by time-variable scattered solar X-rays.  This is grossly
inconsistent with the negligible level of solar X-ray photons mentioned above.
Moreover, BB did not at all explain quantitatively how they arrived at this
scaling factor.  Scaling is complicated, because the cluster fills the entire
field of view, and thus the observed radial X-ray intensity distribution is a
combination of cluster X-rays and cosmic background X-rays.  The cluster profile
is unknown a priori (containing potentially both thermal and excess X-ray
photons).  We must therefore assume that their undertaking was entirely ad hoc.

\subsection{Radial profiles}

BB questioned the reliability of the low-energy calibration of the LECS.  The
low-energy response used by us is based on ray trace simulations.  This is what
is available to the general observer.  BB quote \citet{parmar97} regarding a
discrepancy by a factor of 1.5 at low energies (0.18 and 0.28~keV) between ray
trace simulations and ground measurements.  However, Parmar et al.  suggested
that the discrepancy may be caused by the scattered X-ray photons during the
ground calibration measurements, and that there is yet no convincing
explanation, and that in-flight measurements are needed to resolve this.  Since
neither did BB offer an improved in-flight calibration or ray-trace model, they
essentially use the same calibration data as we did and therefore if our
analysis would contain flaws as a result of this effect, so the same will be
true of theirs.

BB attempt to avoid potential calibration problems (without solving them) by
switching to a radial profile analysis.  First they noted that at low energies,
the FWHM of the instrument is large, in particular in the 0.1--0.3~keV band.
Such a non-uniform resolution is undesirable if radial profiles in different
energy bands are compared.  Therefore, they convolved the radial profile in the
high energy band (0.5--2.2~keV) with a Gaussian of 9.7\arcmin\ FWHM.  Using a
spectral model for the cluster, they then compared this scaled, convolved
0.5--2.2~keV profile with the observed, unconvolved 0.1--0.3~keV band.  The
comparison shows no soft excess, even a small soft X-ray deficit in the center,
and this leads BB to the conclusion that our analysis is wrong.

There are several reasons why this simplified approach by BB is unacceptable.
Specifically:

1.  Convolving the 0.5--2.2~keV image with a Gaussian does not render it
compatible with the resolution of the 0.1--0.3~keV image.  This is mainly due
to the strong, non-Gaussian tails of the instrumental PSF, as we show in the
next section.

2. Degradation of the images by smoothing destroys essential information.

3.  The radial profile in the 0.1--0.3~keV band is also significantly affected
by the low energy response of higher energy photons (32~\% FWHM at 0.28~keV).
Thus, the radial profile in this band is sensitive to spectral variations
as a function of radius.

4.  The effective vignetting corrections to be made are dependent on both energy
and position, and these have been apparently neglected by BB.

5.  The 0.5--2.2~keV band contains the Fe-L complex which can be quite strong in
moderately cool clusters such as A\,2199.  In particular, the strength of this
complex depends on both the amount of central cool gas as well as metallicity.
Since both components vary strongly with position, this biases the 0.5--2.2~keV
flux.

In particular, item 1 is very important for the analysis of the soft X-ray
excess, as we show below.

\section{Radial profiles}

In order to understand what really happens in the analysis of BB, we present here
some simulated radial profiles.  First, we generated a cluster emission profile,
for which we have chosen for demonstration purposes a simple $\beta$-model, with
core radius $r_c$ of 2.6\arcmin\ and $\beta=2/3$, parameters that approximately
describe the structure of the A\,2199 cluster of galaxies.  This image is then
convolved with the monochromatic instrumental PSF of the LECS.  A sufficiently
accurate parameterization for the present purpose, based on the publicly
available calibration files, is given by
\begin{equation}
F(r) = 1 - \left[1 + (x/a)^2\right]^{-b},
\end{equation}
where $F(r)$ is the encircled energy fraction of the instrument.  For energies
of 0.19, 0.277 and 0.93~keV the scale height $a$ is 16.27, 14.70 and
5.30\arcmin, and $b$ is 10.57, 13.32 and 5.12, respectively.  Finally, we
then convolve the high-energy image with a Gaussian for which $F(r) = 1 - \exp
(-r^2/2\sigma^2)$, with $\sigma=4.12$\arcmin, corresponding to the FWHM of
9.7\arcmin\ as used by BB.

\begin{figure}
\plotone{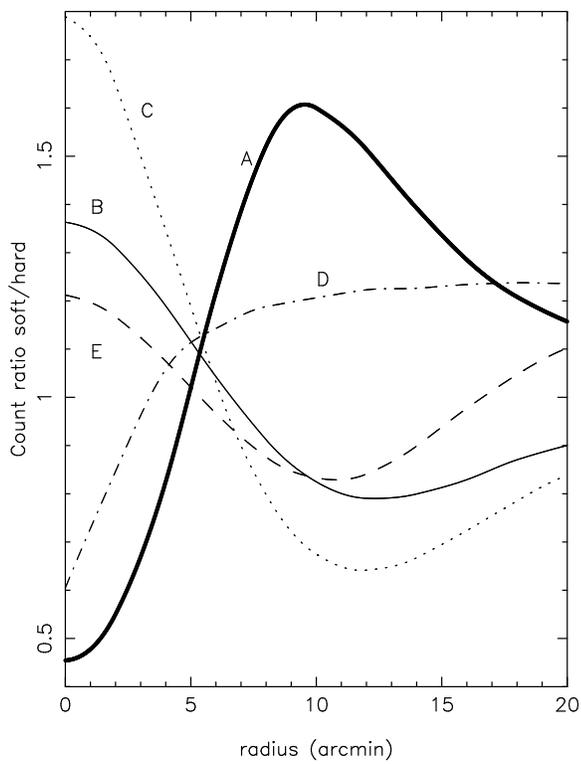}
\caption{Simulated radial profile ratios as described in the text.
 \label{fig1}}
\end{figure}

\placefigure{fig1}

In Figure~\ref{fig1}, we show the results of our analysis.  In all but one case,
we have chosen an effective energy of 0.19~keV for the low energy band (around
the center of the soft 0.1--0.3~keV band of BB), and 0.93~keV for the high
energy band (around the effective center of the 0.5--2.2~keV band of BB).  In
all cases, we plotted the ratio of the soft to hard band, for an equal number of
photons.

Case A in Figure~\ref{fig1} shows the profile ratio for an isothermal cluster,
without convolving the high-energy band with a Gaussian, thus reflecting purely
the instrumental PSF.  The strong drop in the center demonstrates the effect of
the broader PSF at low energies.  It is evident that even for a cluster without
spectral variations, the radial profile ratio varies strongly and is nowhere
equal to unity.

Case B in Figure~\ref{fig1} is similar to case A, but now we convolved the high
energy band with a Gaussian as BB did.  Again, for an isothermal cluster the
ratio varies strongly as a function of $r$, but now with opposite signatures
compared with the previous case.  Apparently, the hard band has been smoothed
too much.  Adopting the approach of BB, a soft flux deficit around 10\arcmin\
would indeed have been inferred, despite the fact that the cluster is
isothermal.  Taking one step further, if BB would have corrected their profile
with a curve like case B, they would have found the soft excess reported by us!

Case C in Figure~\ref{fig1} is similar to case B (i.e., an isothermal cluster
with the BB approach of smoothing the high-energy band), but now for a
low-energy band of 0.277~keV instead of 0.19 keV.  The oversmoothing of the
high-energy band is more evident here.  The comparison of curves B and C shows
that if radial profiles are used for this kind of analysis, the spectral/spatial
energy distribution should be known and modeled appropriately.  This has not
been done by BB.

Case D in Figure~\ref{fig1} simulates the effect of the presence of a cooling
flow or an abundance gradient.  It shows the theoretical hardness ratio in case
the core radius of the hard component is 0.8 times smaller than that of the soft
component.  This mimics, e.g., the case when the Fe-L complex (in the hard band)
is centrally concentrated, because of either an abundance gradient or a cooling
flow.  The profiles have not been convolved with the instrument or the Gaussian.
It also has an $\sim$20~\% excess beyond $\sim$6\arcmin, thereby mimicking
approximately the soft excess as found by us in A~2199.

Finally, case E in Figure~\ref{fig1} corresponds to the case presented in case
D, but now the profiles are convolved with the instrument and the hard band
convolved with the Gaussian.  A comparison of curve E with D shows that the
convolved profile is completely different from the original profile, and
comparing curve E with curve B shows that the differences due to a different
cluster model are partially washed out, thus confirming that the method of BB
tends to destroy information by oversmoothing.

All these effects are completely neglected by BB.  They simply compare their
results with the expected -- scaled -- softness ratio of 1 rather than the
appropriate curve.  That curve can only be obtained by a full spatial/spectral
analysis as, e.g., done by our team.  But then the procedure of BB is obsolete,
since our spectral modeling already gave the correct answers.

\section{Conclusions}

We have shown in this Letter that the analysis method of BB for assessing the
soft excess in A~2199 is inadequate and leads to erroneous conclusions regarding
the presence of a soft excess.

In fact, the method of BB is oversimplified since it neglects the intricacies
associated with the position and energy dependence of the effective area as well
as the spatial/spectral dependence of the LECS PSF.

As a consequence of their inadequate analysis, the results presented by BB are
misleading, in the sense that a simple method which makes use of radial profile
ratios is purported to be better than a full, sophisticated spatial/spectral
analysis, without assessing both methods in a controlled experiment.  As we have
shown here their method leads to unpredictable, erroneous results.

The discussion of BB is also misleading because they claim that the finding of a
soft excess by Kaastra et al.  is due to calibration problems, while BB neither
established this nor used a better LECS calibration.

Finally we have shown that the method of BB is also wrong because of a
misunderstanding of the {\it BeppoSAX} background and corresponding background
subtraction errors.

In conclusion, we have shown in this Letter that our previous results on the
soft X-ray excess in A~2199 are still valid within the uncertainties related to
the spectral and spatial sensitivity of the {\it BeppoSAX} LECS detector.  It is
clear that a further investigation of the presence of the soft excess in
A\,2199, confirming or refuting it, can only come either from a thorough
recalibration of the BeppoSAX instruments or from forthcoming observations with
other satellites with higher sensitivity, such as the {\it XMM-Newton}
satellite.

\acknowledgments

SRON is supported financially by NWO, the Netherlands foundation for Scientific
Research.

%\clearpage

%\clearpage

\clearpage

\end{document}